\documentclass{spataro}

\begin{document}

\markboth{Stefano Spataro for the HADES collaboration}
{$\eta$ MESON RECONSTRUCTION IN PP REACTIONS AT 2.2 GEV WITH HADES}

%
\catchline{}{}{}{}{}
%

\title{$\eta$ MESON RECONSTRUCTION IN PP REACTIONS AT 2.2 GEV WITH HADES}

\author{STEFANO SPATARO}
\author{for the HADES collaboration}
\address{Istituto Nazionale di Fisica Nucleare - Laboratori Nazionali del Sud, Catania, Italy\\ 
II. Physikalisches Institut, Justus-Liebig-Universit\"at, Gie{\ss}en, Germany\\
stefano.spataro@exp2.physik.uni-giessen.de}
\maketitle 
\begin{history}
\received{Day Month Year}
\revised{Day Month Year}
\end{history}
\begin{abstract}
The HADES spectrometer installed at GSI Darmstadt is devoted to study the production of
di-electron pairs from proton, pion and nucleus induced reactions at 1-2 AGeV. 
In pp collisions at 2.2 GeV we have focused mainly on exclusive reconstruction of the $\eta$ meson decays
in the hadronic ($\eta\rightarrow\pi^{+}\pi^{-}\pi^{0}$) and the electromagnetic channels ($\eta\rightarrow e^{+}e^{-}\gamma$). We present analysis techniques and discuss first results on $\eta$ production, with the main focus on comparisons of reconstructed distributions to results obtained by other experiments
and theoretical predictions.
\end{abstract}

\ccode{PACS numbers: 14.40.Aq, 25.40.Ve,13.25.Jx,13.40.Hq}

\section{The Experiment}	

The HADES spectrometer, which is described in more details in the contribution of Ref.~\refcite{Jur}, is a highly selective tool specially suited for the study of the high energy dielectron decay channels in nucleus-nucleus collisions around 1-2 A GeV, as well as in proton and pion induced reactions. In order to provide absolute measurements of invariant mass distributions a good knowledge of the dielectron reconstruction efficiency is mandatory, which can be achieved by means of well known calibration reactions, as the $pp\rightarrow pp\eta$ channel.\\
A proton-proton run was done on January 2004 at 2.2GeV, to measure the $pp\rightarrow pp\eta$ reaction through an exclusive reconstruction of the hadronic ($\eta\rightarrow\pi^{+}\pi^{-}\pi^{0}$) and the dielectron ($\eta\rightarrow e^{+}e^{-}\gamma$) Dalitz decays, whose branching ratios are known. This fact allows to use the $pp\rightarrow pp\eta$ channel as a calibration reaction for the dielectron identification, in order to normalize the dielectron yields in theavy ions experiments.
The study of the $\eta$ meson is of particular interest for several reasons. In the vicinity of the threshold the production of the $\eta$ meson was studied extensively and the inclusive cross sections are well known\cite{BAL01}, while at higher energies the uncertainties are larger. Moreover the observed electron pair invariant mass distribution for the $\eta$ Dalitz decay branch can be used to check contributions from other sources (such as $\pi^{0}$ Dalitz, combinatorial background, etc.) as well as the lepton trigger efficiency. In addition, if both protons and both leptons are detected, the reconstruction of this decay mode is kinematically complete without requiring an additional photon detector, and a missing mass technique can be applied. This would allow to separate the $\eta$  Dalitz from all other sources and to map the effective acceptance of the spectrometer.
\section{Time-of-flight measurements}
In HADES, however, the main algorithm for particle identification requires a combination of momentum and time-of-flight measurements. In the pp experiment it was not possible to use a START detector as a time reference, because of the high number of secondaries produced. This means that there is no common start time reference for tracks in the same event, but in general the data acquisition is started by the fastest particle which crosses the time-of-flight wall. In this case we measure the difference in time with respect to the fastest particle, instead of the real time-of-flight. Thus it was necessary to develop an algorithm to reconstruct the start time of the reaction, and make particle identification possible.\\
The used algorithm relies on identifying one particle in the event and calculate its theoretical time-of-flight: from the assumed mass and the measured momentum we calculate the time-of-flight of the particle, thus the offset to the real start time of the reaction. The first kind of identification comes from the Cherenkov detector, which is hadron blind and can select electron/positron tracks. If a lepton candidate is not found, a negative charged track in the same event is assumed as a negative pion. Finally, we recalculate the time-of-flight of the other particles inside the same event using the extrapolated reaction time. In this way we obtained an average time resolution of 340 ps
and an efficiency of about 92\% for events with a lepton, and a resolution of
440 ps and an efficiency of about 93\% for events with a negative pion.\\
After the start time reconstruction it is possible to use the recalculated time-of-flight in order to identify particles, for exclusive analysis of $\eta$ decay channels.

\section{Exclusive $\eta$ reconstruction}
\begin{figure}[t]
\begin{center}
\includegraphics[width=0.42\textwidth]{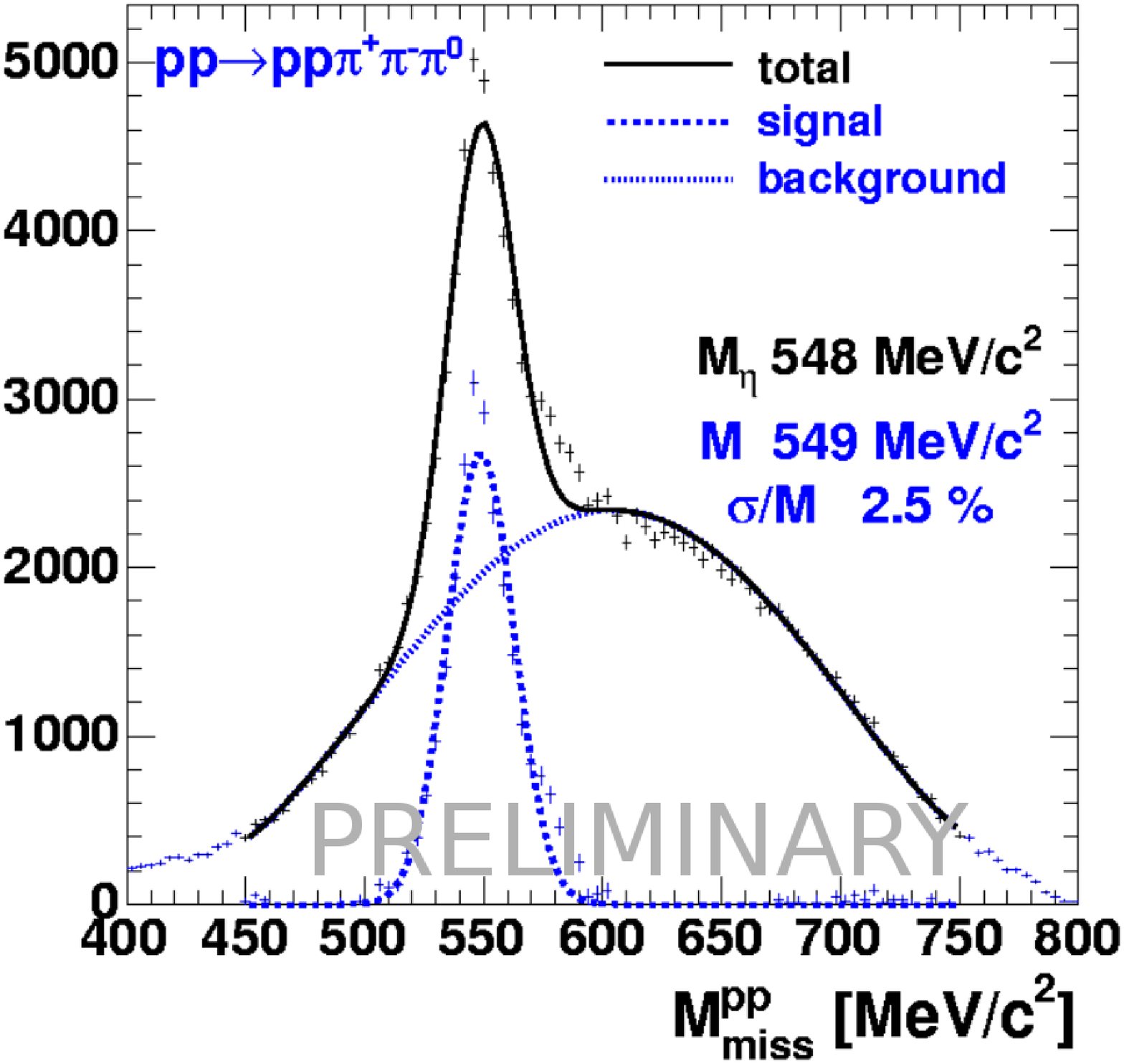}
\includegraphics[width=0.42\textwidth]{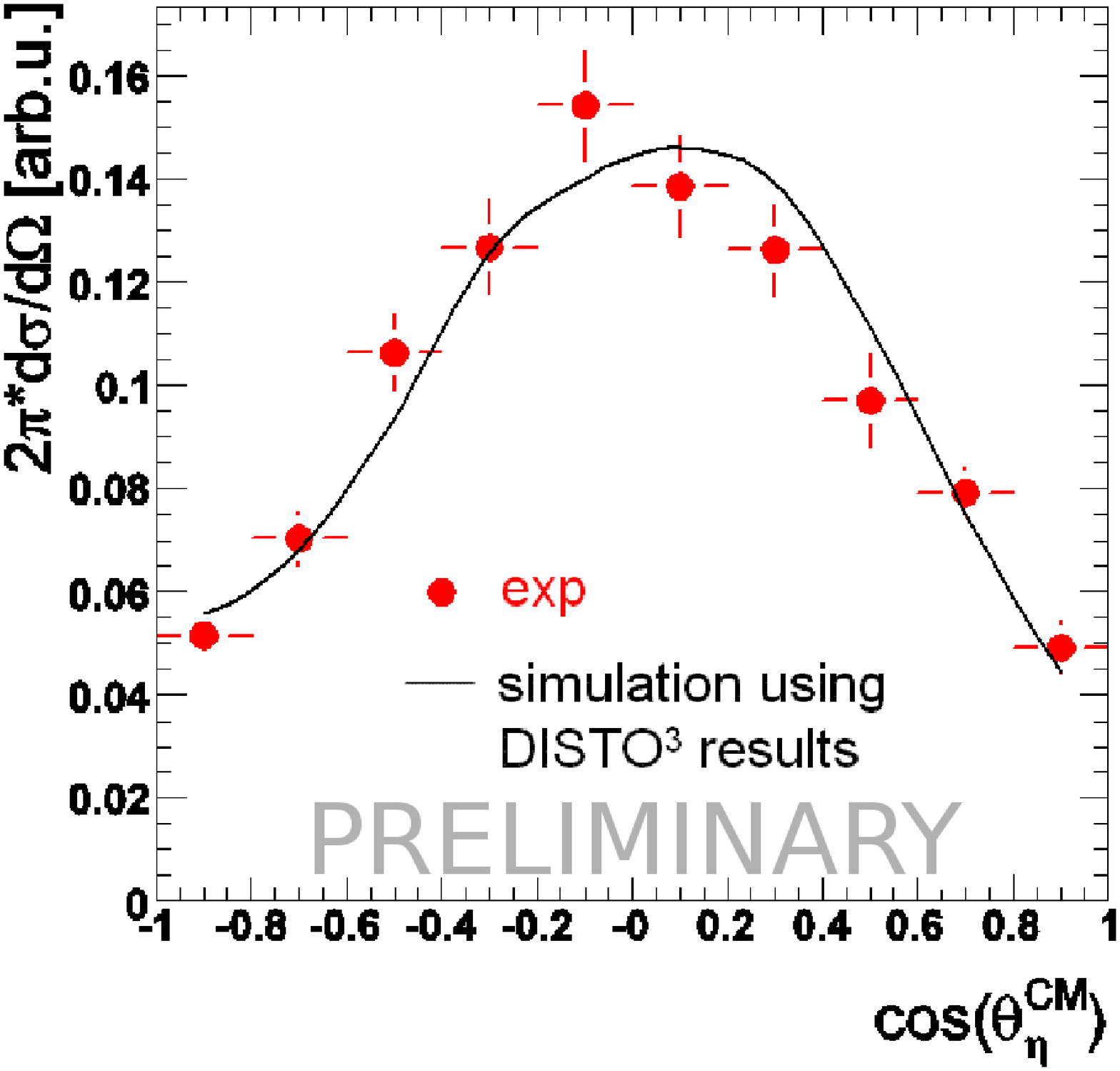}
\caption[]{{\it Left}: Proton-proton missing mass plot for the hadronic decay channel, after a kinematic refit procedure. {\it Right}: Distribution for the $\eta$ center-of-mass polar angle for $\eta\rightarrow\pi^{+}\pi^{-}\pi^{0}$ events, without correction for acceptance end efficiency.}
\label{etahad_plot}
\end{center}
\end{figure}

\begin{figure}[t]
\begin{center}
\includegraphics[width=0.42\textwidth]{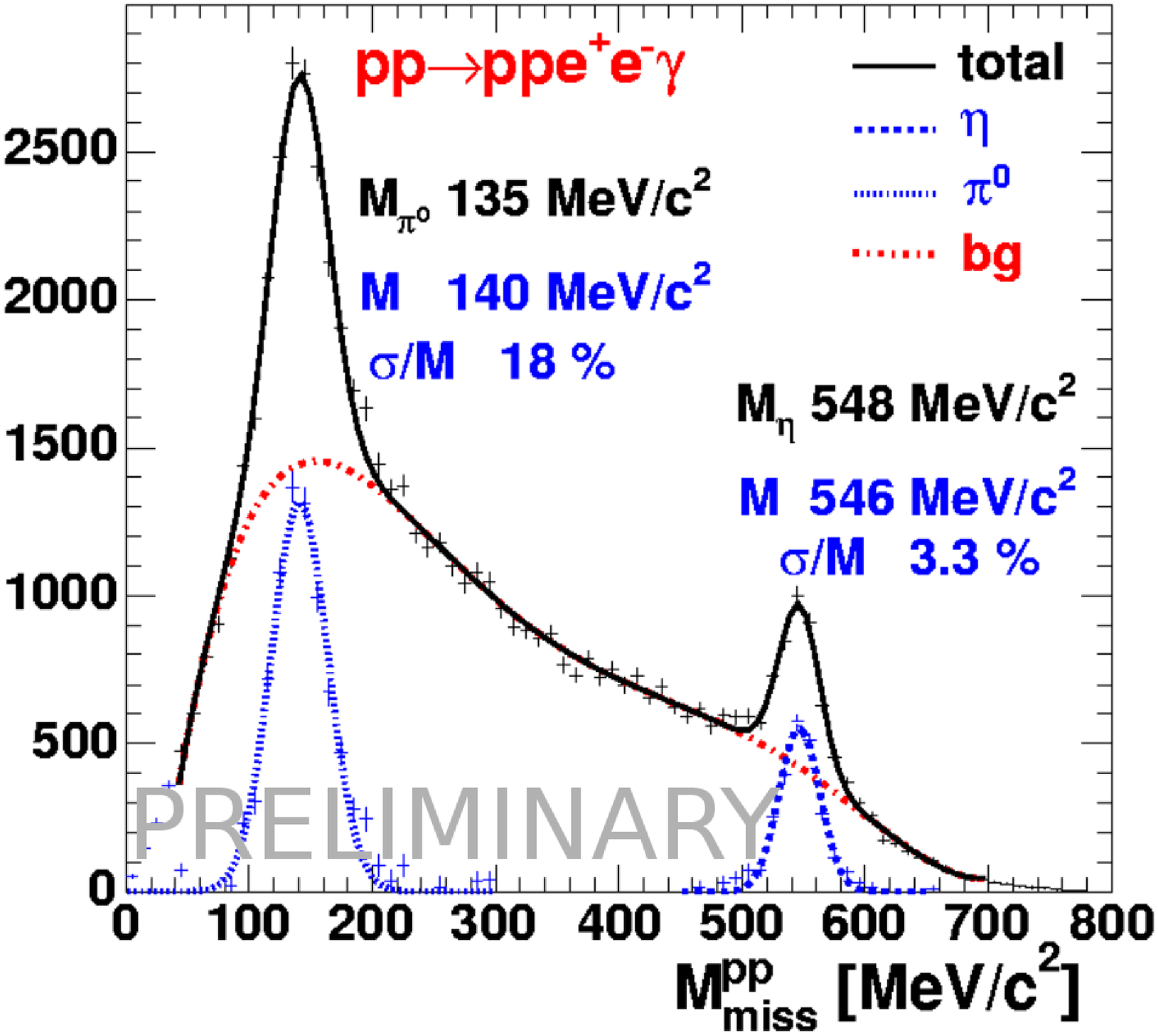}
\includegraphics[width=0.42\textwidth,height=130pt]{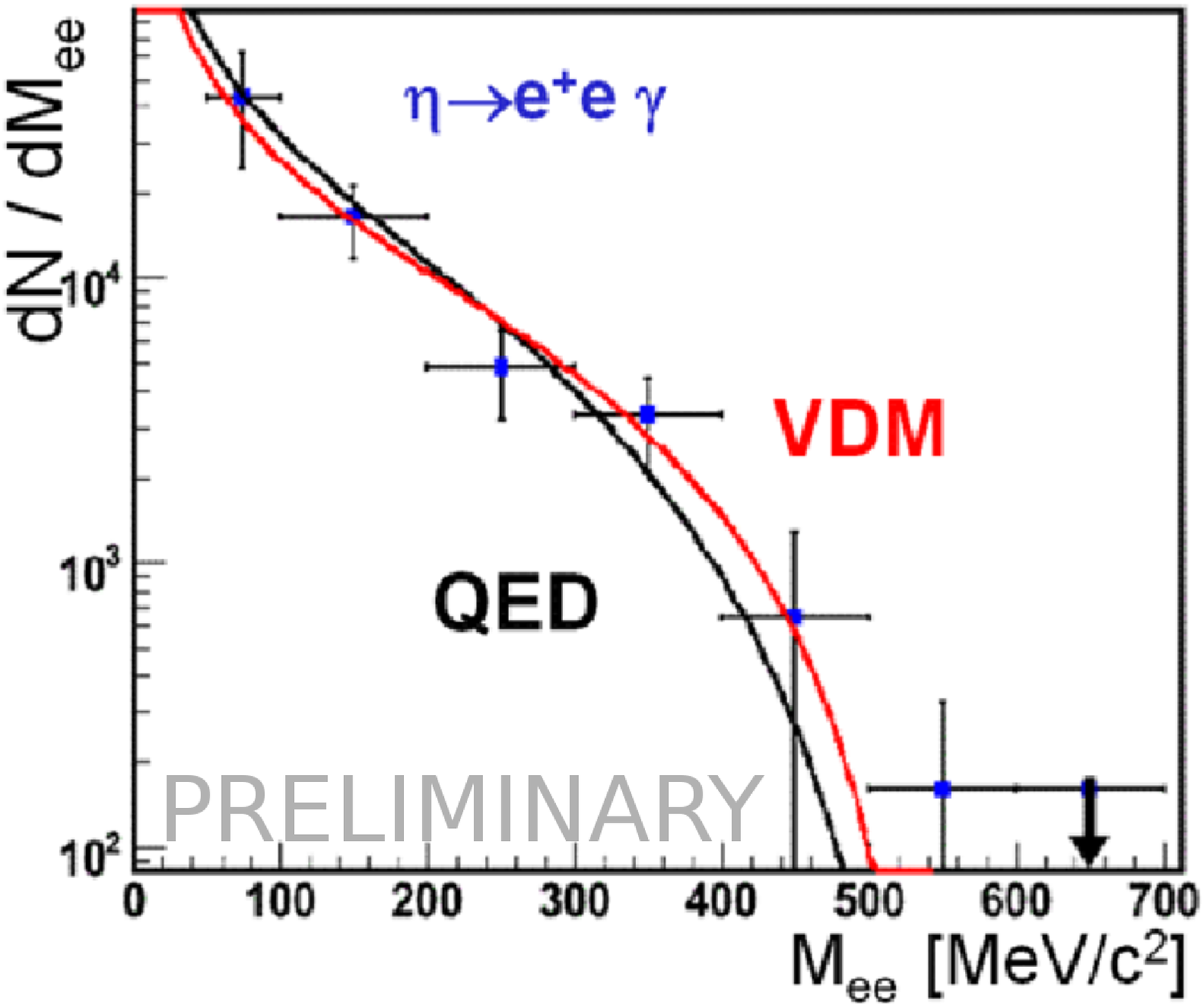}
\caption[]{{\it Left}: Proton-proton missing mass plot for electromagnetic $\eta$ meson decay, after a kinematic refit procedure. {\it Right}: Invariant mass distribution of dielectron pairs coming from $\eta$ decay, after the conversion rejection, in comparison with simulation by using  QED or VDM $\eta$ form factor.}
\label{etadal_plot}
\end{center}
\end{figure}

The main goal of the p+p experiment was to verify the dielectron reconstruction efficiency needed for a correct interpretation of the HI data, by means of the well known $\eta$ meson decays into hadronic\cite{BAL04} and electromagnetic channels\cite{PDG}.
In the simulation we assumed resonant production of the meson via $pp\rightarrow pN(1535)\rightarrow pp\eta$, with the Dalitz matrix element taken from Ref.~\refcite{BAL04}. For the eta hadronic decay the matrix element from Ref.~\refcite{CRB} has been implemented, and for the $\eta$ Dalitz decay the vector dominance model (VDM) was assumed. In addition a full cocktail of reaction channels based on measured proton-proton cross sections\cite{Land} was generated. Simulated events were processed through the HADES GEANT code and reconstructed with the same analysis program used for the experimental data.
Reconstruction of the $pp \rightarrow pp \eta \rightarrow pp \pi^{+}\pi^{-}\pi^{0}$ reaction started with charged particle identification (PID) based on momentum and reconstructed time of flight. After PID a kinematical condition was used on events with two protons and two charged pions: by imposing a cut of $3\sigma$ around the $\pi^{0}$ region in the four-particle missing mass we selected the $\eta\rightarrow \pi^{+}\pi^{-}\pi^{0}$ reaction, $\sigma$  being determined from a fit to the missing $\pi^{0}$ peak. The missing-mass resolution was further improved with a kinematical refit of the full  event\cite{Anar}. Finally, the $\eta$ mesons were identified in the two-proton missing-mass distribution shown in the left plot of Fig. \ref{etahad_plot}. A prominent $\eta$ peak, centered at the expected position, is clearly visible on top of a non-resonant three-pion background.
By fitting the distribution by means of a gaussian plus a background  function, we obtain the value of the $\eta$ mass with a resolution of 2.5\%, and evaluate the hadronic production yield.
Moreover, by fitting the pp missing mass peak in various $cos\theta^{CM}_{\eta}$ slices\cite{Fro}, the angular distribution on the $\eta$ meson emission has been evaluated and compared to simulation, by using the anisotropic angular model verified by the DISTO collaboration\cite{BAL04}(right plot of Fig. \ref{etahad_plot}). The two distributions are in agreement within error bars, showing that the hadron efficiency is understood in the covered phase space region.
The reconstruction of the $pp \rightarrow pp \eta \rightarrow ppe^{+}e^{-}\gamma$ decay was done in a similar way. As a first step, events with two proton and two electron tracks were selected, thus the $ppe^{+}e^{-}\gamma$  reaction channel was identified by imposing a $3\sigma$ selection on the four-particle missing mass around the missing photon peak. The left side of Fig. \ref{etadal_plot} shows the proton-proton missing mass distribution, obtained after a kinematical refit was applied to the $ppe^{+}e^{-}\gamma$ events with a constraint on a missing photon. The prominent peaks centered at 140 and 546 $MeV/c^{2}$ correspond to the $\pi^{0}$ and $\eta$ mesons respectively.
The istribution is dominated by events $\eta\rightarrow\gamma\gamma$ (branching ratio $\sim40\%$) where one of the decay photons produces a conversion pair. However, as the simulation also shows, it can be completely removed by conditions on the quality of the track reconstruction and the pair opening angle. The right plot of Fig. \ref{etadal_plot} shows the invariant mass distribution of dielectron pairs coming from the $\eta$ decay after the conversion rejection\cite{Spr}, in comparison with simulation by using QED (point-like interaction) and VDM ($\eta$ form factor) theoretical models. The agreement demostrates our good understanding of the dielectron invariant mass spectrum, and that the form factor measurement is feasible with the HADES spectrometer.\\
Finally we compare the ratio R of the $\eta$ yields\footnote{Yields before the kinematic refit procedure were considered, to avoid systematic effects.} reconstructed in the hadronic and electromagnetic decays (including conversion contribution) to the predictions of the simulation with the following results: $R_{exp}= 15.3 \pm 1.8 (stat)$ and $R_{sim} = 15.6 \pm 0.9 (stat)$. The agreement in between the two values shows how the simulation well reproduces the experimental yields, and it confirms the global good understanding of the response of the HADES spectrometer.
\section*{Acknowledgments}
This work has been supported by GA CR 202/00/1668 and GA AS CR IAA1048304 (Czech Republic), KBN 5P03B 140 20 (Poland), BMBF (Germany), INFN (Italy), CNRS/IN2P3 (France), MCYT FPA2000-2041-C02-02 and XUGA PGIDT02PXIC20605PN (Spain) and INTAS ref. Nr. 03-51-3208


\begin{thebibliography}{0}    
\bibitem{Jur} J. Pietraszko {\it et al.}, this conference.
\bibitem{BAL01} F. Balestra {\it et al.}, {\it Phys. Rev. C} {\bf 63}, 024004 (1991).
\bibitem{BAL04} DISTO Collab. {\it F. Balestra et al.}, {\it Phys. Rev. C} {\bf 69}, 064003 (2004).
\bibitem{PDG} Particle Data Group, {\it Phys. Lett. B} {\bf 592} (2004).
\bibitem{CRB} Crystal Barrel Collab. {\it C. Amsler et al.}, {\it Phys. Lett. B} {\bf 346}, 203 (1995).
\bibitem{Land} Landolt-B\"ornstein, {\it New Series I/12b}.
\bibitem{Anar} A. Rustamov, {\it PhD Thesis}, TU Darmstadt (2006).
\bibitem{Ste} S. Spataro, {\it PhD Thesis}, Universit\'a degli Studi di Catania, Italy (2006).
\bibitem{Fro} I. Fr\"ohlich, {\it IVth Int. Conf. on Quarks and Nuclear Physics (QNP06)}, Madrid (2006).
\bibitem{Spr} B. Spruck, {\it PhD Thesis}, Justus-Liebig-Universit\"at, Gie{\ss}en, Germany.
\end{thebibliography}
\end{document}